\documentclass[%
 aip,
 jmp,%
 amsmath,amssymb,
 preprint,%
prl]{revtex4-1}

\usepackage{graphicx}% Include figure files
\usepackage{dcolumn}% Align table columns on decimal point
\usepackage{bm}% bold math
\usepackage[section]{placeins}
\begin{document}
%\newcounter{subfigure}
%\preprint{AIP/123-QED}

\title[Elucidating plasma dynamics in HW turbulence by information geometry]
      {Elucidating plasma dynamics in Hasegawa-Wakatani turbulence by information geometry} % Force line breaks with \\
%\thanks{Footnote to title of article.}

\author{Johan Anderson}%
\email{anderson.johan@gmail.com.}
\affiliation{ 
Department of Earth and Space Sciences, Chalmers University of Technology, 
SE-412 96 G\"{o}teborg, Sweden
}%

\author{Eun-jin Kim}
\affiliation{ 
School of Mathematics and Statistics, University of Sheffield, Sheffield S3 7RH, United Kingdom 
}%

\author{Bogdan Hnat}
\affiliation{ 
Department of Physics, University of Warwick, Coventry, CV4 7AL, United Kingdom
}%

\author{Tariq Rafiq}
\affiliation{
Department of Mechanical Engineering and Mechanics, Lehigh University, Bethlehem, PA 18015, USA
}%

\date{\today}% It is always \today, today,
             %  but any date may be explicitly specified

\begin{abstract}
The impact of adiabatic electrons on drift-wave turbulence, modelled by the Hasegawa-Wakatani equations,
is studied using information length. Information length is a novel theoretical method for measuring distances
between statistical states represented by different probability distribution functions (PDFs) along the path of a system. 
Specifically, the time-dependent PDFs of turbulent fluctuations for a given adiabatic index $A$ is computed. The changes in fluctuation statistics are then quantified in time by using information length. The numerical results provide time traces exhibiting 
intermittent plasma dynamics, and such behaviour is identified by a rapid change in the information length.
The effects of $A$ are discussed.
\end{abstract}

                             % Classification Scheme.
\maketitle

%%%%%%%%%%%%%%%%%%%%%%%%%%%%%%%%%%%%%%%%%%%%%%%%%%%%%%%%%%%%%%%%%%%%%%%%%%%%%%%%

\section{Introduction}
\label{sec:intro}
Turbulence is ubiquitous in nature, and is of fundamental importance to many space and laboratory plasma systems.
In magnetically confined (MC) plasmas, turbulent fluctuations in plasma potential, density and temperature cause elevated transport compared to classical predictions \cite{horton1999}. In MC plasmas, a large radial gradient is present resulting in a strong anisotropy between parallel and perpendicular length scales. The non-adiabatic electron response is a crucial component of their dynamics.

Investigations of this turbulent transport is an outstanding topic in fusion plasma research. Qualitative understanding can be obtained by studying reduced models, such as Hasegawa - Mima (HM) or the Hasegawa - Wakatani (HW) equations \cite{horton1999, hm1977, hm1978, hmk1979, hw1983, hw1986, dewhurst2008, dewhurst2009, dueck2013, anderson2017}. The reduced models are more amenable to analytical elucidations as well as detailed numerical predictions. The HM and HW equations include fixed electron pressure gradient, which provides a driving mechanism for drift waves. These waves become unstable on the gyro-radius scale in the presence of parallel electron resistivity, present in the HW model. Non-linear interactions of large amplitude drift-waves leads to a turbulent state. This turbulence is quasi-2D, that is the energy condensates on 
large scales, hence the system evolves from random perturbations on small scales to an ordered state dominated by large-scale structures where the average size of a structure depends on the adiabatic parameter A. The ability to associate certain transport characteristics with particular physics of the model elucidates experimental results, as well as predictions from quantitative but more complex counterparts such as gyrokinetic models. In particular, HW model \cite{hmk1979, hw1983, hw1986, dewhurst2008, dewhurst2009, dueck2013} is in an intermediate regime between adiabatic and hydrodynamic electrons, allowing the electrons to dynamically and self-consistently determine the relation between the density and the electrostatic potential through the turbulence. 

Recently, the need to investigate large scale transport events such as bursts, streamers and blobs have been recognized  \cite{hi1996, zweben2007, politzer2000, beyer2000, drake1988, antar2001, carreras1996}. These intermittent events are characterized by a bursty temporal structure and radial coherence. The Probability Distribution Functions (PDFs) of fluxes associated with these events have elevated tails compared to a Gaussian distribution, which can be a manifestation of large events or coherent structures mediating transport \cite{politzer2000}. This statistical intermittency is quantified by higher order cumulants of the PDFs (e.g. skewness and kurtosis) \cite{anderson2010, kim2008}. However, a key dynamical feature of magnetically confined plasma includes a different kind of structure, which is radially localized while extended in the poloidal direction \cite{anderson2010, kim2008, anderson2014, anderson2018}. These are known as zonal flows. Zonal flows are generated by the small scale turbulence and may act in a self-regulating manner and govern the saturation of the drift waves \cite{DiamondEA2005, ConnorEA2004, ItohEA2006, ConnorMartin2007}.  Theoretical models have previously successfully predicted the functional form of the PDF tail for the electrostatic fluctuations described by drift turbulence models \cite{anderson2010, kim2008}. 

Understanding the time evolution of in and out of equilibrium systems is one of the major goals in statistical mechanics. One of the possible options is to work with the different metrics for the thermodynamical length \cite{weinhold1975, rupeiner1979, schlogl1985, diosi1996, crooks2007, feng2009} and the information length \cite{nicholson2015, kim2017, kim20172, kim2019,Geometry} which is a generalisation to non-equilibrium systems. The thermodynamic length endows a phase space with a Riemannian metric, thus allowing one to measure the distance that a system travels between thermodynamic equilibrium states. Thus it constitutes a geometric methodology to understand stochastic processes involved in order-disorder transition.  Interestingly, one possible choice is to use the PDFs to construct the Fisher information metric\cite{nicholson2015, kim2017}. This metric gives a novel methodology to measure distance in statistical space. Thus it is possible to assess the difference in the dynamics between two time points. When a PDF continuously changes with time, the information length measures the total number of different statistical states that a system passes through in time \cite{nicholson2015, kim2017, kim20172, kim2019,Geometry}.

In this work, we investigate quasi-stationary time series of the electrostatic potential and corresponding vorticity (poloidally averaged and sampled at different radial points) from the HW simulations. We compute the information length $\mathcal{L}$, show examples of the time evolution of the PDFs and discuss implications and possible events in the time traces. Here, the information length quantifies the differences between different statistical states of the system during its evolution.

The paper is organized as follows. The HW model is described in Section \ref{sec:HW} and the statistical analysis with interpretation methods are explained in Section \ref{sec:INF}. The results are presented for electrostatic potential, vorticity and flux in Section \ref{sec:potential}, \ref{sec:vort} and \ref{sec:flux}, respectively. The paper is concluded with a discussion in Section \ref{discuss}. 

%%%%%%%%%%%%%%%%%%%%%%%%%%%%%%%%%%%%%%%%%%%%%%%%%%%%%%%%%%%%%%%%%%%%%%%%%%%%%%%%

\section{Hasegawa - Wakatani model}
\label{sec:HW}
As noted previously, the Hasegawa-Wakatani system of equations (HW) provide a reduced model that has frequently been employed to study transport processes in magnetically confined (MCF) plasma \cite{hmk1979}. There exist quantitatively better and increasingly complex models, however studies of HW offers an alternative where sampling of long time series data where the time evolution of the PDFs can be monitored is enabled. These numerical simulations based on HW capture the key elements of MCF plasma dynamics: drift instability due to non-adiabatic electron response, onset of drift turbulence and the self-organisation of plasma into zonal flows. 

The HW describes low frequency ($\omega \ll \omega_{ci}$, where $\omega_{ci}$ is the ion gyro frequency) fluctuations of the density $n$ and the electrostatic potential $\phi$, in the presence of the constant background density gradient, parallel electron resistivity and for a small ion-electron temperature ratio ($T_i/T_e \ll 1$). In the presence of axisymmetric zonal flows with poloidal wave number $m=0$, which do not contribute to the parallel currents, the HW is in the quasi two dimensional form:
\begin{eqnarray}
\frac{\partial n}{\partial t} &=&-\kappa \frac{\partial \phi}{\partial y}+A (\tilde{\phi}-\tilde{n})+[n,\phi]+D \nabla^2 n,
\label{HW1}\\
\frac{\partial}{\partial t} \nabla^2\phi &=&A (\tilde{\phi}-\tilde{n})+[\nabla^2\phi,\phi]+\mu \nabla^2 (\nabla^2 \phi), 
\label{HW2}
\end{eqnarray}
where total fluctuating fields $n$ and $\phi$ consist of turbulent parts, $\tilde{n}$, $\tilde{\phi}$, and
zonal fluctuations $\langle n \rangle$ and $\langle \phi \rangle$. That is, $n=\tilde{n}+\langle n \rangle$ and 
$\phi=\tilde{\phi}+\langle \phi \rangle$. The poloidal averages are denoted by the angular brackets $\langle \dots \rangle$, which in the slab model simply indicate integration along the poloidal line at a given radial location:
\begin{equation}
 \langle f \rangle= \frac{1}{L_y} \int_0^{L_y} f dy.
\label{fsa}
\end{equation}

The nonlinear advection terms are expressed as Poisson brackets
$[A,B] = \partial A/\partial x .\partial B/ \partial y - \partial A/\partial y .\partial B/\partial x$. In both equations physical quantities have been normalized using $e\phi/T_e \to \phi$, $n/n_0 \to n$, $\omega_{ci} t \to t$ and $(x,y)/\rho_s \to (x,y)$. Standard notation is used for other quantities: $T_e$ is the electron temperature, $\omega_{ci}$ is the ion gyrofrequency, and $\rho_s = \sqrt{m_i T_e}/eB$ is the hybrid Larmor radius. Dissipation terms of the form $\nabla^2 \phi$ are added to the equations for numerical stability, where $D$ and $\mu$ are dissipation coefficients. The dissipation coefficients have physical interpretations where $D$ is identified with the cross-field ambipolar diffusion and $\mu$ is the ion perpendicular viscosity. The $x$ and $y$ directions are identified with the radial and poloidal directions in a tokamak, respectively, and the magnetic field is assumed to point in the $z$ direction. We assume that $\kappa = -\partial \ln(n_0)/\partial x$ determines the background density profile $n_0(x)$. In Eqs. (\ref{HW1})-(\ref{HW2}), there is a free parameter $A$ that controls the strength of the resistive coupling between $n$ and $\phi$ through the parallel current,
\begin{eqnarray}
A = \frac{T_e k^2}{n_0 e^2 \eta \omega_{ci}}
\end{eqnarray}
where $\eta$ is electron resistivity. This is called the adiabaticity parameter, $A$ determines the degree to which electrons can move rapidly along the magnetic field lines and establish a perturbed Boltzmann density response. There exist to limits of interest for the HW system and it approaches different regimes in the limits of $A \to 0$ and $A \to \infty$. When $A \to \infty$ and $n \to \phi$ (that is density fluctuations become enslaved to the electrostatic potential fluctuations) the HW become identical to a one field indirectly forced Charney-Hasegawa-Mima equation \cite{hm1977,hm1978}. In the opposite limit $A \to 0$ the equation becomes equivalent to the incompressible Euler equation in 2D. A non-zero $A$ yields a growth rate causing a random stirring that prevents the vorticity from decaying to zero as is the usual case with the unforced Navier-Stokes. Note that there are both stable, unstable waves and non-modal solutions present in the system. 

The equations \eqref{HW1} and \eqref{HW2} were solved numerically on a square grid of size $L=40$ (units of $\rho_s$) with spatial resolution of $256\times256$ grid points. It is well known that the final stage of the HW evolution is dominated by zonal flows. In real laboratory plasmas and in 3D simulations, the poloidal damping of zonal flows enforces a certain level of energy equipartition between turbulence and zonal flows. Equipartition in our simulations is enforced by applying the following algorithm. We modify the decomposition of fluctuations so that each fluctuating quantity $Q=\tilde{Q}+\gamma \langle Q \rangle$ ($Q=n, \phi$), that is we multiply each poloidally averaged zonal flow component by factor $\gamma$. In each step of the simulation we monitor the kinetic energy of zonal flows
\begin{equation}
 \langle E \rangle_K \equiv \frac{1}{2} \int \left( \frac{\partial \langle \phi \rangle}{\partial x} \right)^2 dV,
\label{ZKE}
\end{equation}
and its turbulent counterpart, given by
\begin{equation}
E_K \equiv \frac{1}{2} \int \left( \nabla \phi \right)^2 dV.
\label{TKE}
\end{equation}
If at any time step $\langle E \rangle_K > E_K$, we set $\gamma=E_K / \langle E \rangle_K$, otherwise we set $\gamma=1.0$
\cite{dewhurst2009}.

We performed three simulations varying the adiabaticity $A=0.25$, $1.0$ and $2.0$, which capture three distinct dynamical regimes for the HW system. We examine poloidally averaged potential and vorticity, where the data is collected in ten evenly spaced radial locations starting from location $x=20$ and ending at $x=200$.
%%%%%%%%%%%%%%%%%%%%%%%%%%%%%%%%%%%%%%%%%%%%%%%%%%%%%%%%%%%%%%%%%%%%%%%%%%%

\section{Time Dependent PDFs and Information Length}
\label{sec:INF}
For simplicity, let us consider a stochastic variable $x$ with time-dependent PDF $p(x,t)$. When we know the control parameters $\lambda^i$ 
that determine the PDF, we can calculate the Fisher-Information metric $g_{ij}$ 
\begin{equation}
g_{ij} = \int dx p(x,t) \frac{\partial \log p(x,t)}{\partial \lambda^i} \frac{\partial \log p(x,t)}{\partial \lambda^j}.
\label{gij}
\end{equation}
Noe that the function $p(x,t)$ determines the probability of the system to be in state $x$ at time $t$.
Based on this metric tensor, we can write down the information length \cite{nicholson2015, kim2017, kim20172, kim2019} as
% \cite{weinhold1975, rupeiner1979, schlogl1985, diosi1996, crooks2007, feng2009, nicholson2015} 
\begin{equation}
\mathcal{L} = \int_0^{\tau} dt \sqrt{\frac{d \lambda^i}{dt} g_{ij} \frac{d \lambda^j}{dt}}.
\label{lth}
\end{equation}
The distance in Eq. (\ref{lth}) is measured by the difference between consecutive PDFs where the difference in PDFs gives a measure of the statistical distance. The evolution of a system can then be envisioned as the trajectory in the probability space where the distance/metric at different times is provided by the statistical distance. \\

In general, even when we do not know control parameters that govern PDFs, we can calculate the information length \cite{nicholson2015, kim2017, kim20172, kim2019,Geometry}
by following the main two steps, computing the dynamic time unit $\tau(t)$ and the total time in this unit. The dynamic time is the characteristic time scale over which $p(x,t)$ temporally changes on average at time $t$. The second step involves computation of the total elapsed time in units of this $\tau(t)$.
Specifically, the dynamic time $\tau (t)$ is related to the second moment $\mathcal{E}$ (as can be inferred from combining Eqs. (\ref{lth}) and (\ref{gij})) can be computed by,
\begin{equation}
\mathcal{E} = \frac{1}{|\tau (t)|} = \int dx \frac{1}{p(x,t)} \left(\frac{\partial p(x,t)}{\partial t} \right)^2,
\label{etau}
\end{equation}
and quantifies the correlation time over which the (dimensionless) information changes.
The information length $\mathcal{L}(t)$ then follows
\cite{nicholson2015},
\begin{equation}
\mathcal{L}(t) = \int_0^t \frac{ds}{\tau (s)} = \int_0^t ds \sqrt{\int dx \frac{1}{p(x,s)} \left(\frac{\partial p(x,s)}{\partial s}\right)^2}.
\label{Int}
\end{equation}
Note that none of the PDFs is identically zero, which would lead to a singularity in (\ref{Int}). The information length is dimensionless and represents the total different number of states between the initial and final times, $0$ and $t$ respectively, and establishes a distance between the initial and final PDFs in the statistical space. The simplest case elucidating this principle is a Gaussian distribution, statistically distinguishable states are determined by the standard deviation, which increases with the level of fluctuations. If two PDFs have the same standard deviation while differing in peak positions by less than one standard deviation, these two PDFs and the two processes are then statistically indistinguishable. The information length was shown to be proportional to the time integral of the square root of the infinitesimal relative entropy \cite{Geometry}.

In the analysis of HW we focus on the time traces (averaged in the $y$-direction) at four equidistant radial points located at $x=40, 80, 120, 160$. The original simulation data sets are down-sampled to more amenable data set that consists of typically $5\times 10^5$ entries. Thus, we employ the same analysis method for the electrostatic potential, vorticity and flux for all values of the adiabaticity parameter (A) from $0.25$, $1.00$ and $2.00$. We present all cases in turn comparing the results for each value of adiabaticity index, starting with the potentials, and PDFs and then finally the information length.

%%%%%%%%%%%%%%%%%%%%%%%%%%%%%%%%%%%%%%%%%%%%%%%%%%%%%%%%%%%%%%%%%%%%%%%%%%%%%%
\section{Results for potential with varying adiabaticity}
\label{sec:potential}
In this section the results of the analysis is presented. We start with a few observations from the simulations and the effect of the adiabaticity index $A$.
%Figure 1
%Figure phi time trace A=0.25, 1.00 and 2.00
\begin{figure}[ht]
{\vspace{2mm}
\includegraphics[width=10cm, height=7cm]{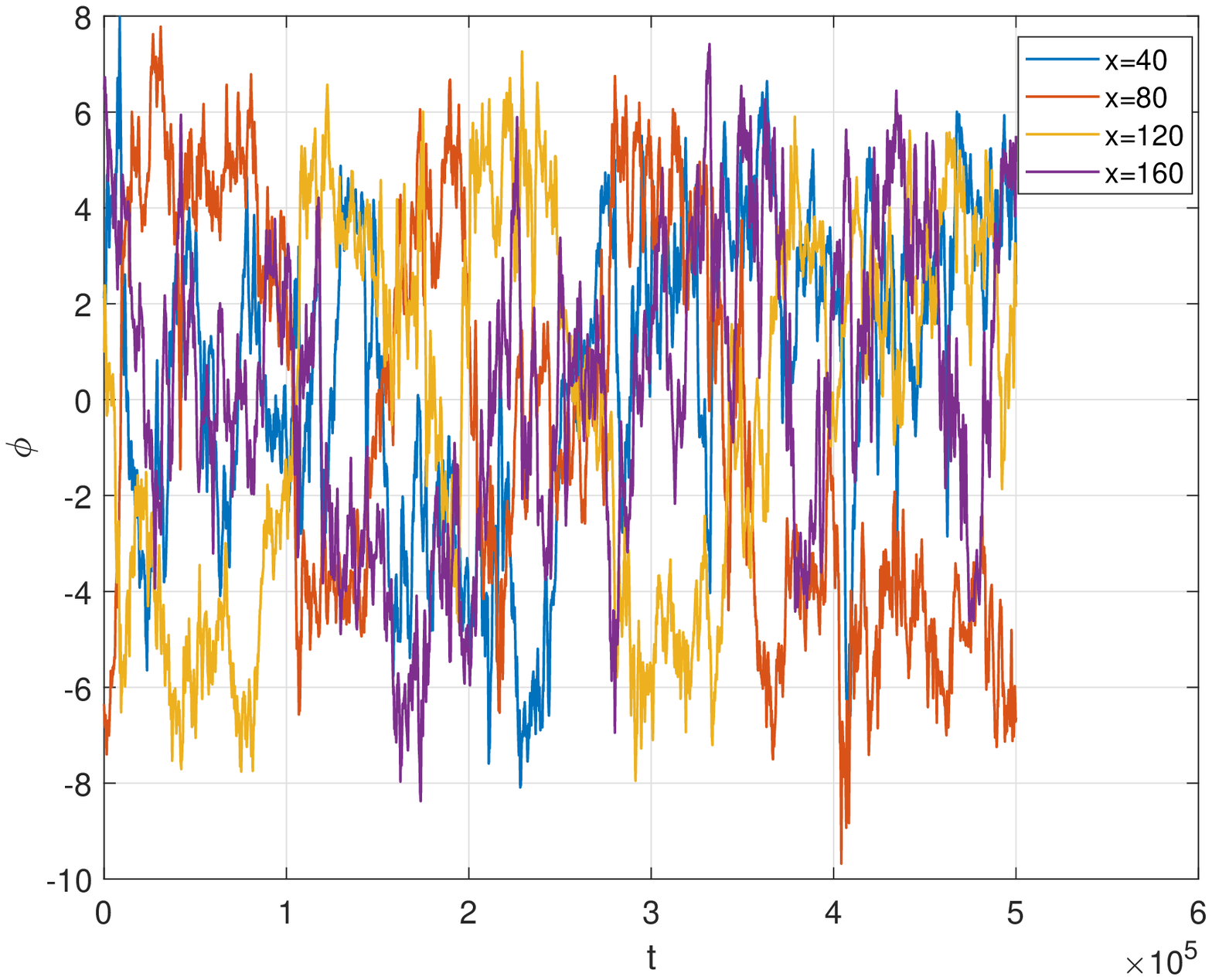}}
{\includegraphics[width=10cm, height=7cm]{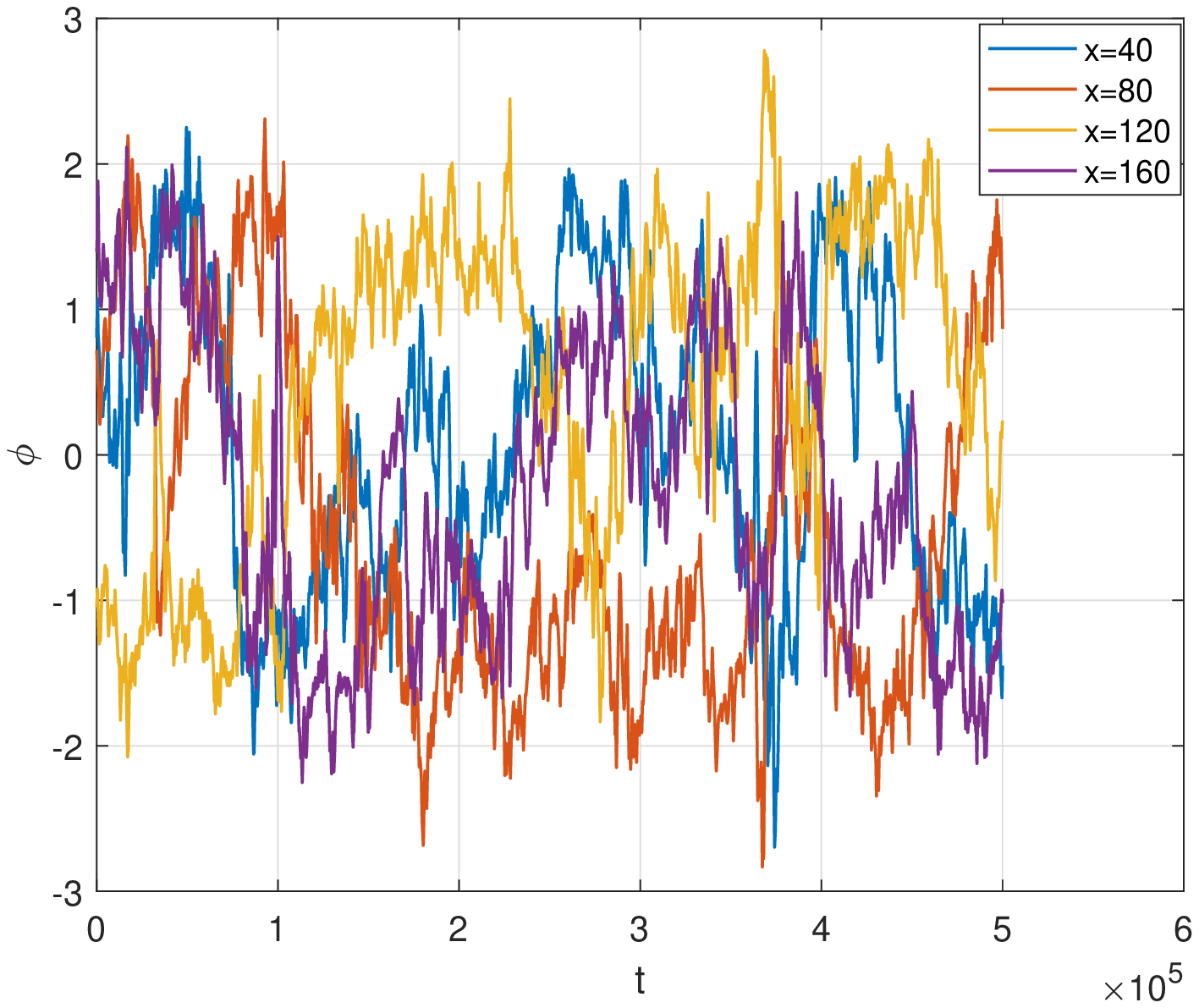}}
{\includegraphics[width=10cm, height=7cm]{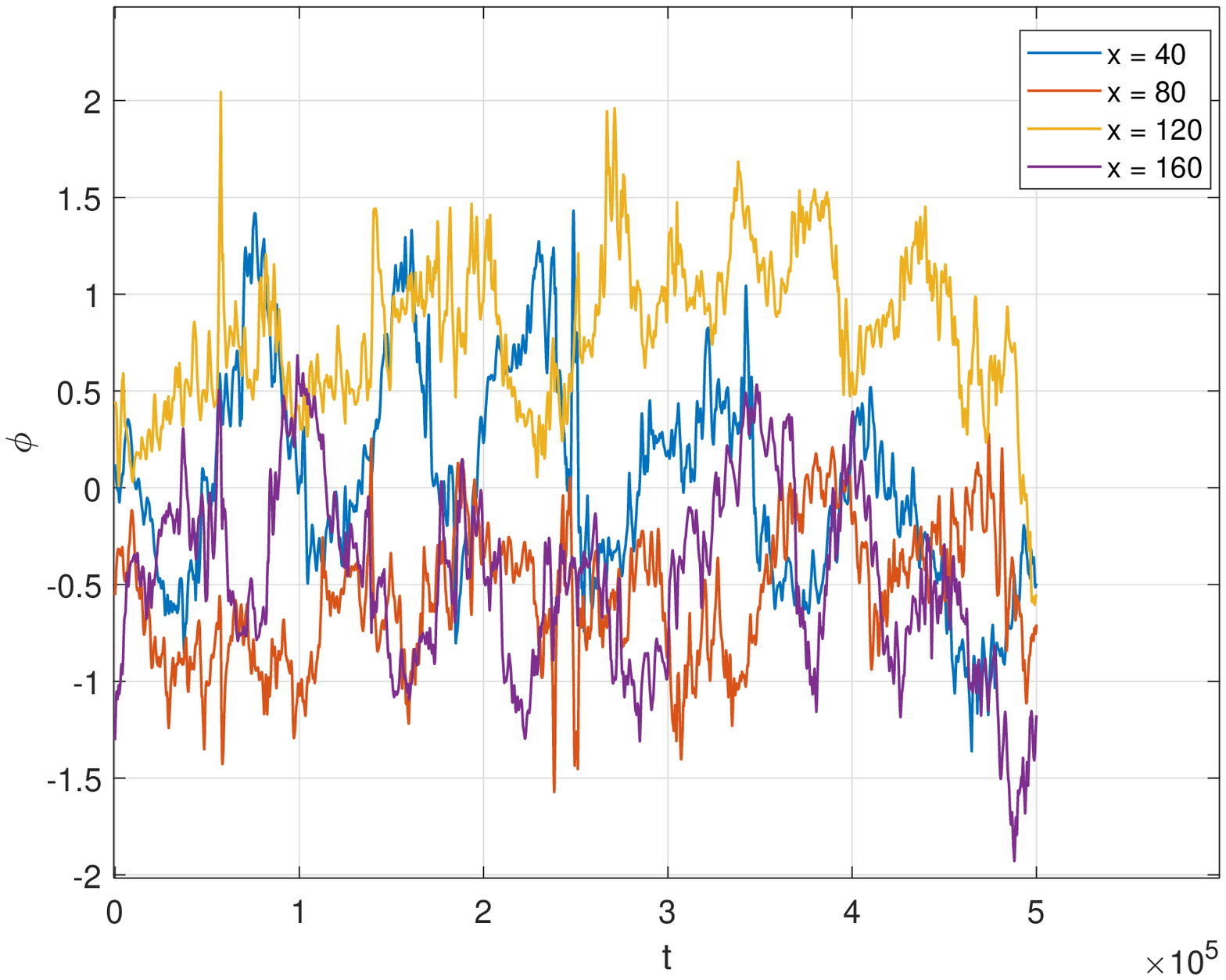}}
\caption{
The time trace of potential at the radial positions (40, 80, 120, 160) for $A = 0.25$ (top), $1.00$ (middle) and $2.00$ (bottom).
}
\label{f:fsaphi}
\end{figure}
In Figure (\ref{f:fsaphi}) the time traces of the potentials at the radial positions $(40, 80, 120, 160)$ are presented for $A = 0.25, 1.00$ and $2.00$. These are the evolution of the mean value of the potentials averaged over the poloidal direction. We note that there are some significant differences between the three regimes described by the variation in adiabaticity. For small adiabaticity ($A=0.25$) there are several superimposed fluctuations on different time scales clearly visible whereas for larger adiabaticity the longer time scale fluctuations seem to be of importance.

Interestingly, also the fluctuation levels decrease as the adiabaticity is increased corresponding to the different regimes where small adiabaticity corresponds to Euler like dynamics and high adiabaticity HM limit, respectively. Specifically, for $A=0.25$, the time traces of the potentials at the different radial positions span a similar interval between $-10$ and $8$ with the similar statistical property (e.g. standard deviation) (results not shown.). As $A$ increases, the behaviour at the different $x$ becomes noticeable. For instance, for $A=2$, the absolute values of $\phi$ for the red and purple lines ($x=80$ and $120$) are mostly smaller than those for the blue and yellow lines ($x=40$ and $x=120$). This indicates that there is a difference in the dynamics between the edge ($x=40$ and $x=160$) and the core ($x=80$ and $x=120$). The dynamics will be quantified by computing PDFs at the corresponding radial points comparing the time evolution of these PDFs.

%Figure 2
%Figure PDFs A=0.25
\begin{figure}[ht]
{\vspace{2mm}
\includegraphics[width=10cm, height=7cm]{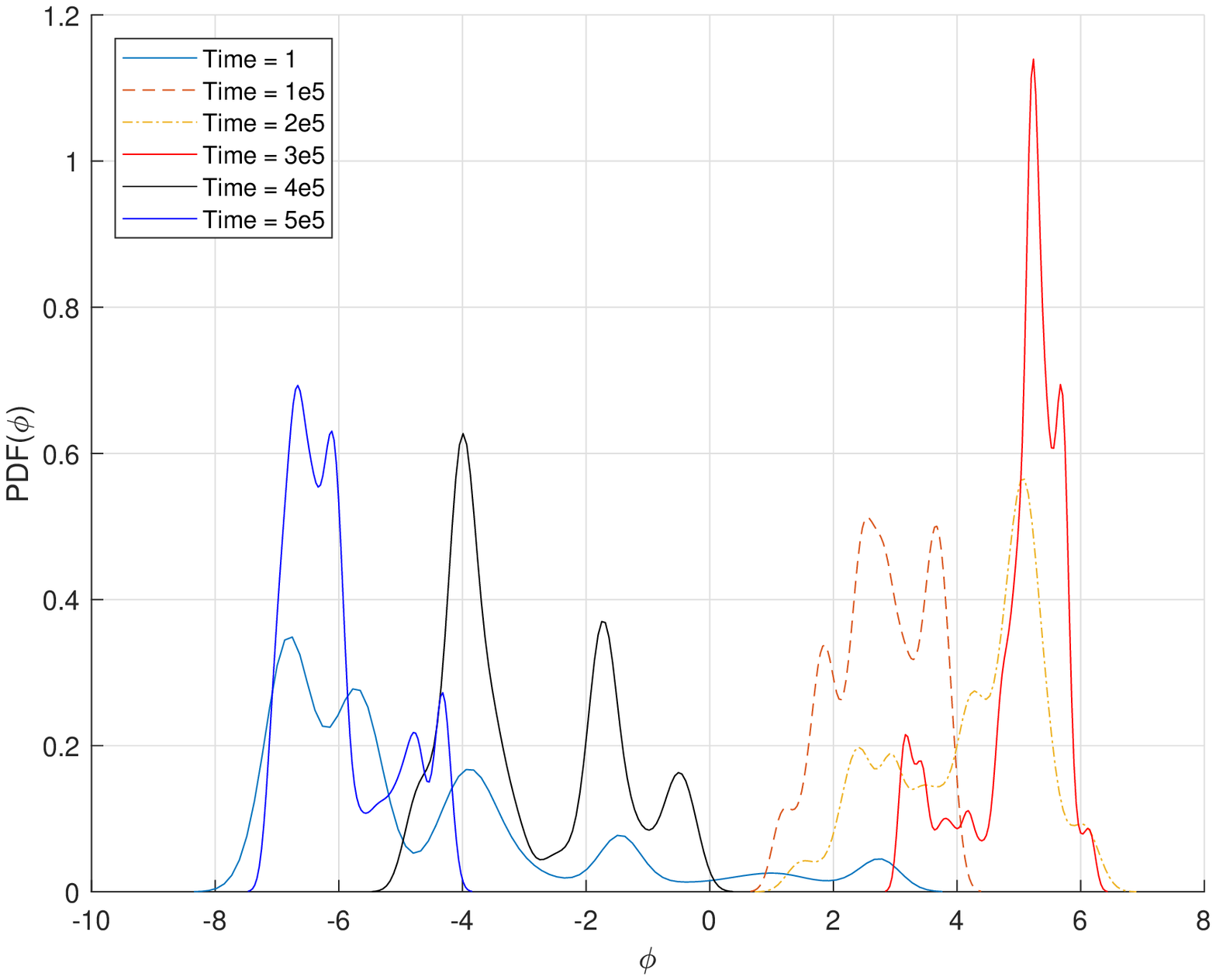}
}{\includegraphics[width=10cm, height=7cm]{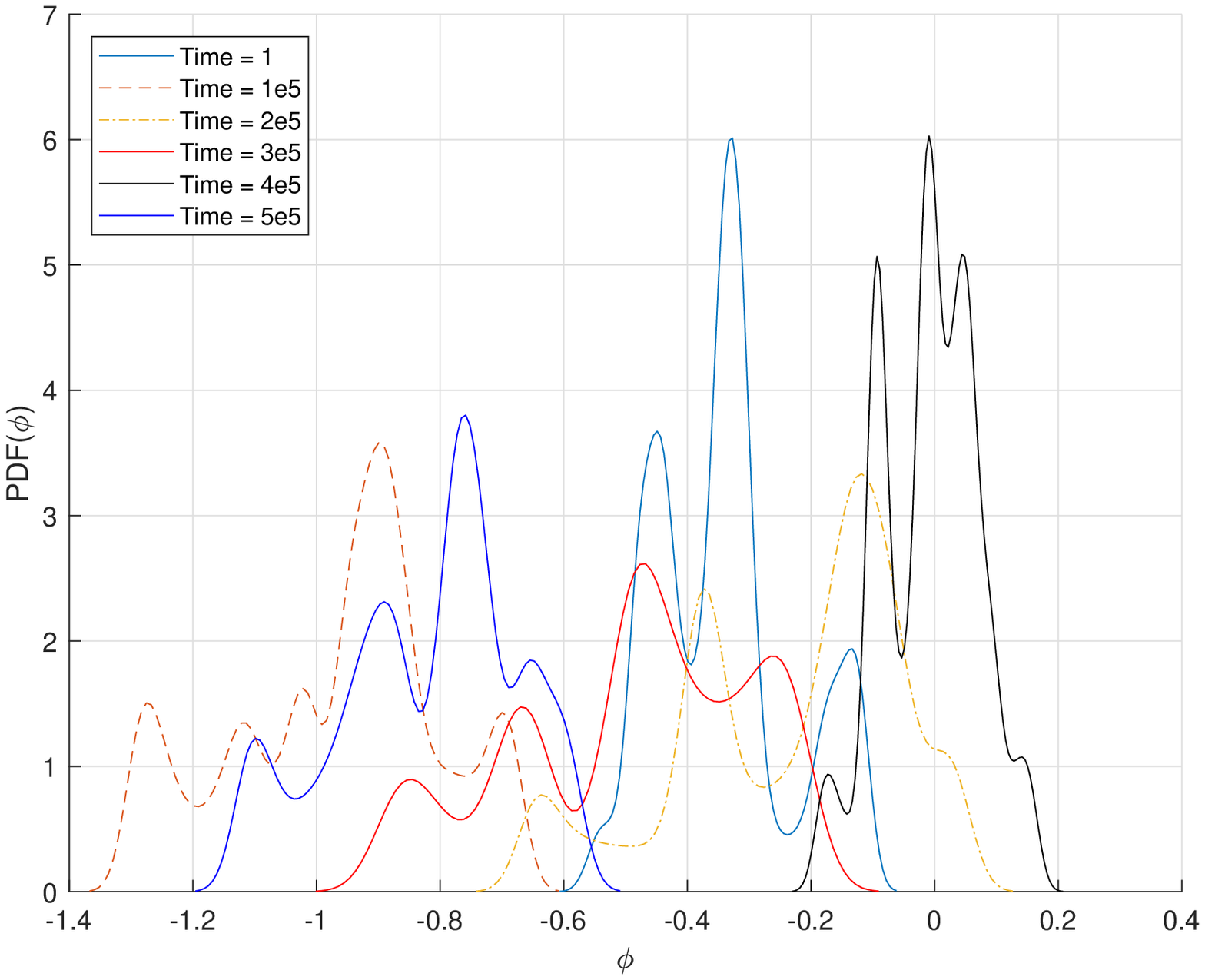}}
\caption{
The time evolution of PDFs of the potential at x = 80 for adiabaticity $A = 0.25$ (top) and $A = 2.00$ (bottom).
}
\label{f:PDFs_x4}
\end{figure}
An example of the time evolution of the PDFs are shown in Figure (\ref{f:PDFs_x4}). The PDFs are constructed by using 10000 consecutive values in time of the potential $\phi$, the large fluctuations are clearly visible in the rather wide (large standard deviation) PDFs however also the large difference in mean is clearly visible. The information length assesses the distance between distributions at different instances in time. Note that, although the distribution functions can be constructed from linear combinations of the non-linear invariants of the starting HW equations this is only on average and not the quasi-stationary time-dependent PDFs. 
%Figure 3
%Figure Information length
\begin{figure}[ht]
{\vspace{2mm}
\includegraphics[width=10cm, height=7cm]{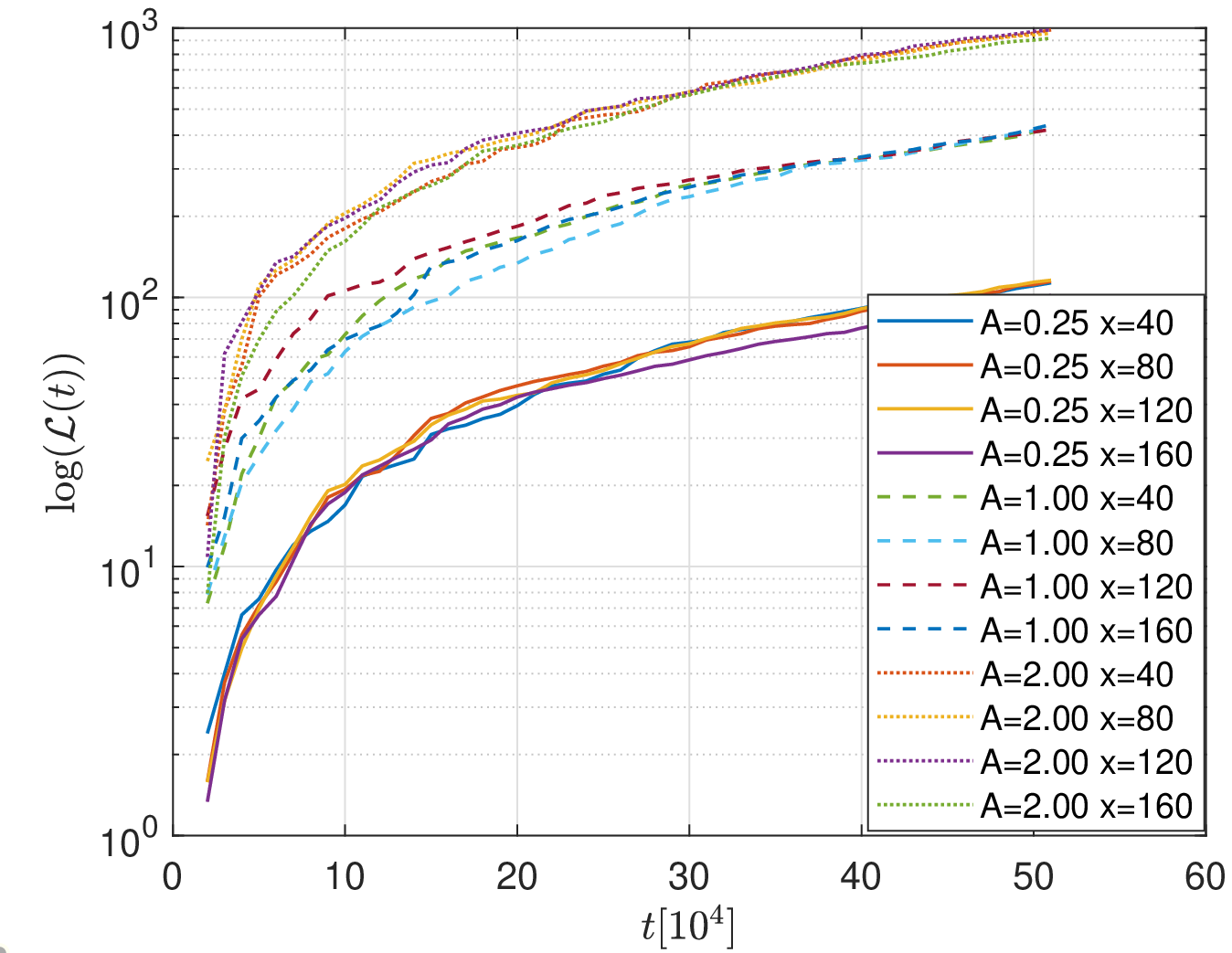}
}
\caption{
The logarithm of the information length computed using Eq. (\ref{Int}), for the four radial points and $A = 0.25, 1.00$ and $2.00$.
}
\label{f:fsaphi_int}
\end{figure}
Figure \ref{f:fsaphi_int} presents the information length (log scale) computed for the radial positions (40, 80, 120, 160) with adiabaticity ($A$) as a parameter. Note that the information length is monotonically increasing in time although at different rates; the information length saturates at a constant (stationary) value only when the PDFs are independent of time. Interestingly, the fast fluctuation visible for $A = 0.25$ has a smaller information distance between the PDFs compared to the smaller fluctuation found for $A = 2.00$. This is because of a much higher fluctuation level and thus much higher uncertainty for $A=0.25$ than for $A=2.00$. We also mote that, the PDFs are dependent on the bin sizes and a sensitivity study using bin sizes of $500, 1000, 5000$ and $10000$ was considered where the resulting PDFs were rather stable for the larger bin size. The information length is naturally dependent on the resolution however small variation for all tested cases were found. Moreover, it was noted that in Figure \ref{f:fsaphi} for $A = 0.25$ there is a variation on a longer time scale and halfway through the simulation the potential at $x=40$ and $=120$ flips from the negative side to the positive and vice versa during a short time. However, this is not immediately visible in Figure \ref{f:fsaphi_int}, thus a more in-depth assessment is called for.
%Figure 4
%Figure Information length A = 0.25
\begin{figure}[ht]
{\vspace{2mm}
\includegraphics[width=10cm, height=7cm]{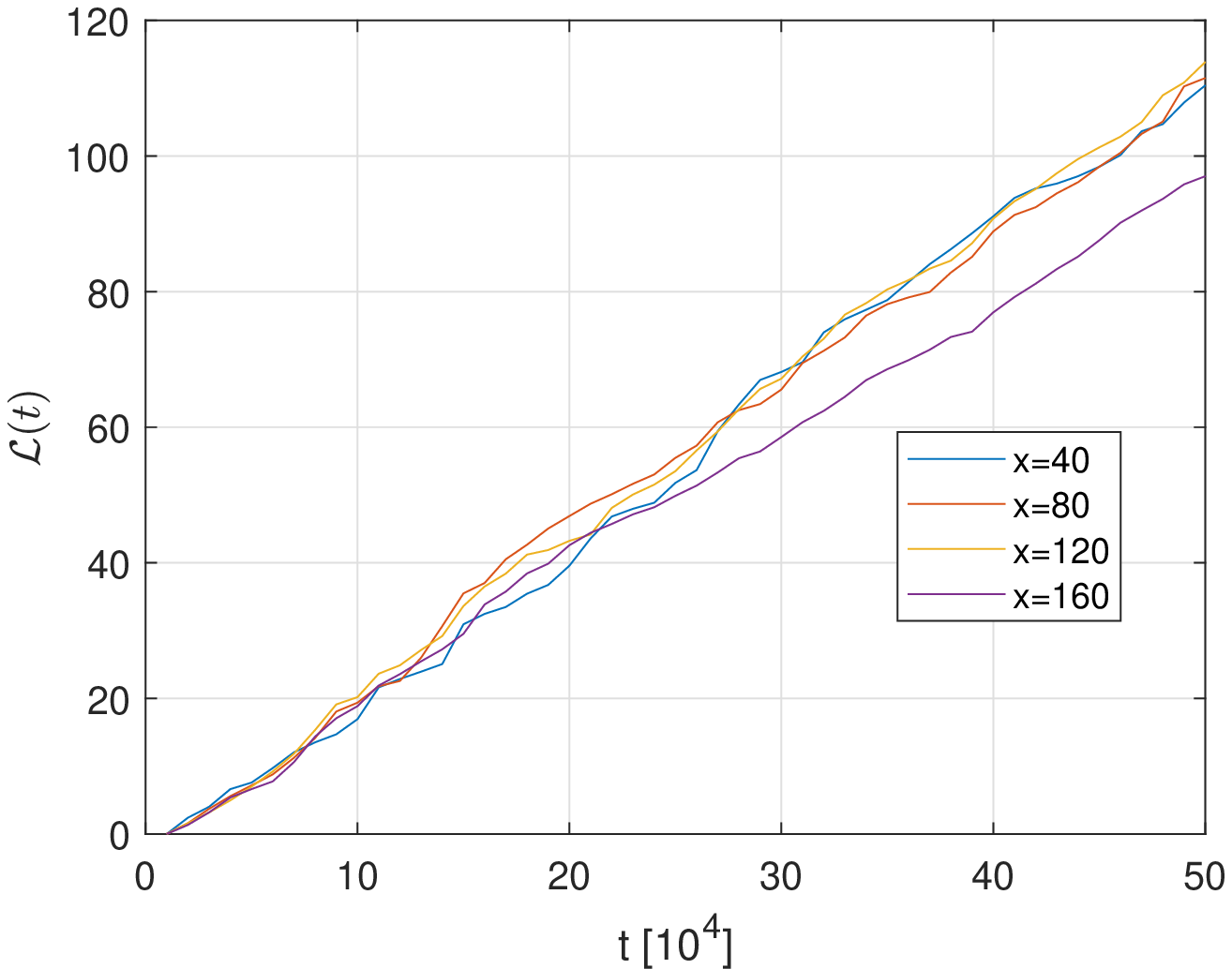}
}
\caption{
The information length for small adiabatic index $(A = 0.25)$, for the four radial points $x = 40, 80, 120$, and $160$.
}
\label{f:fsaphi_int025}
\end{figure}
The information length is displayed in Figure \ref{f:fsaphi_int025} where there is a linear increase in the information without significant change approximately where the change in the potential occurs. However, looking at the individual contributions at each time instance gives additional clues. We note that the expression in Eq. (\ref{etau}) for the dynamic length is positive definite and thus the integral sum will be monotonically increasing. 
\begin{figure}[ht]
{\vspace{2mm}
\includegraphics[width=10cm, height=7cm]{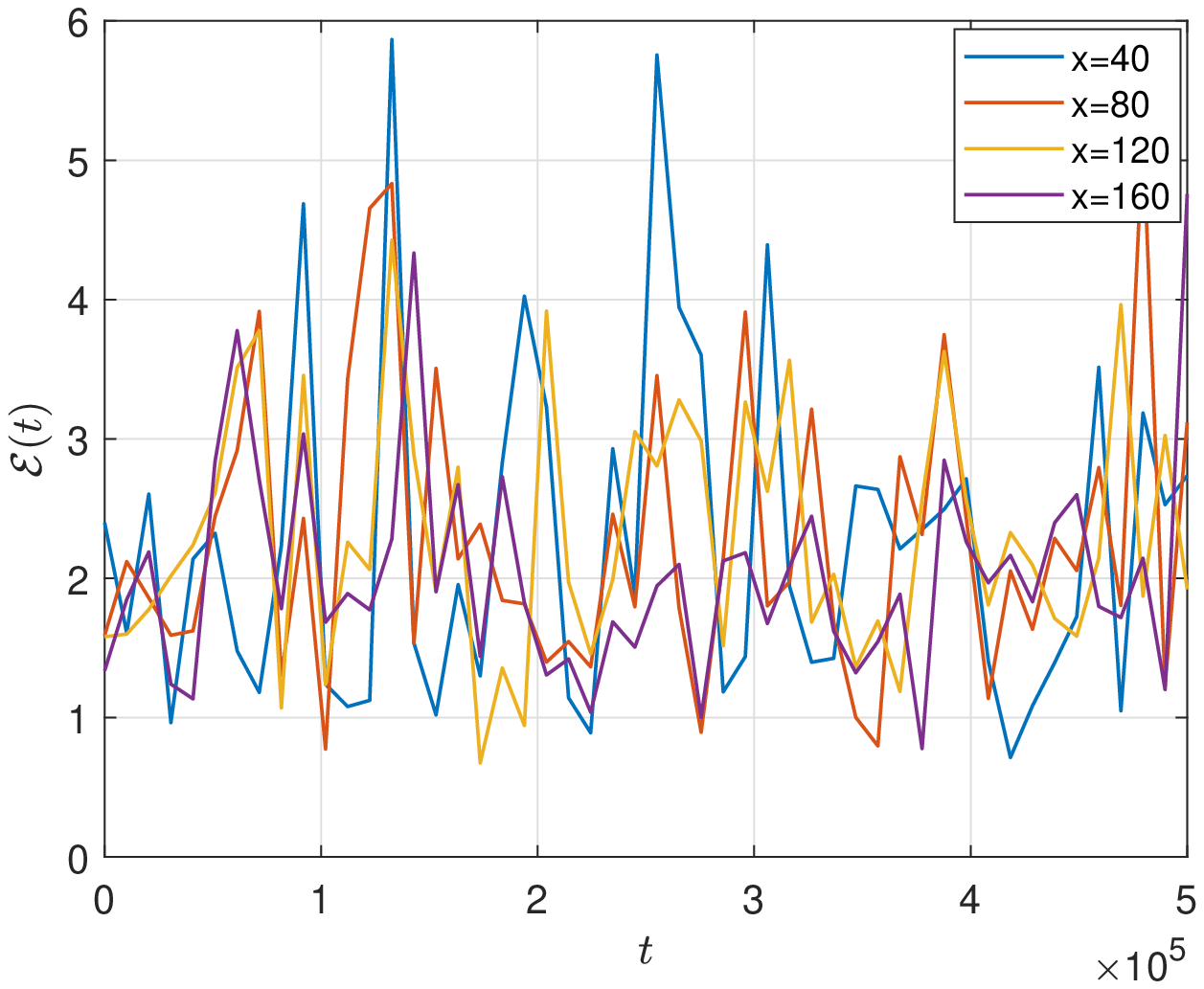} \includegraphics[width=10cm, height=7cm]{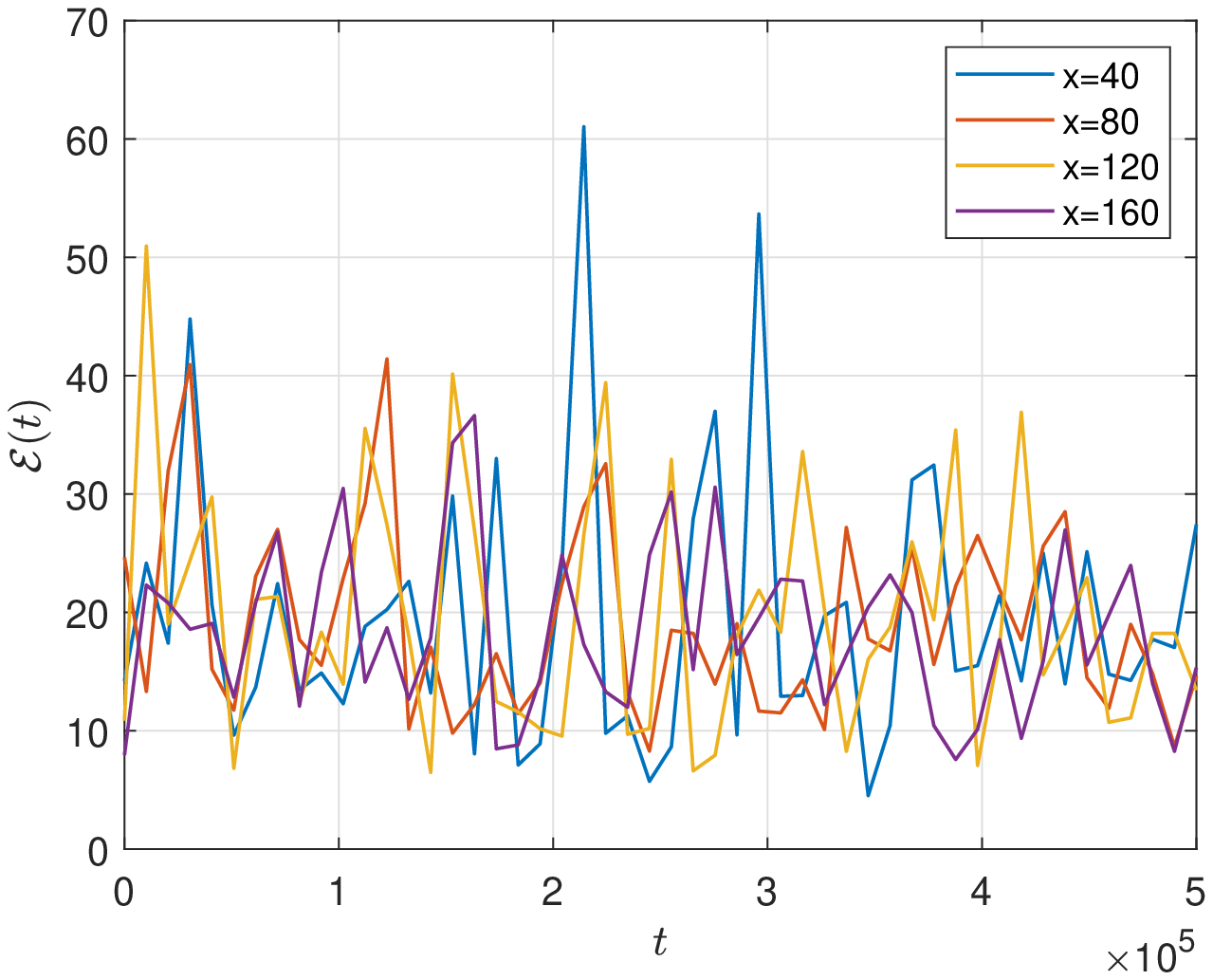}
}
\caption{
The dynamic time at small $A = 0.25$ (top) and large $A=2.00$ (bottom) for the four radial points $x = 40, 80, 120$, and $160$.
}
\label{f:tau_dyn}
\end{figure}
Figure~\ref{f:tau_dyn} shows ${\cal E} = 1/\tau^2$ at small $A = 0.25$ (top) and large $A=2.00$ (bottom) for the four radial points $x = 40, 80, 120$, and $160$ computed according to Eq. (\ref{etau}). The instances with large values of ${\cal E}$ coincides as expected with the sudden change in information length. In the top panel of Figure~\ref{f:tau_dyn} for $A=0.25$, we observe large variation of ${\cal E}$ over time for all $x=20,80,120, 160$. However, the blue line ($x=40$) shows the largest variation, reaching up ${\cal E} \sim 6$. This means that for $x=40$, the potential PDF undergoes a sudden change on a small time scale $\tau \propto {\cal E}^{-1/2}$. This behaviour is to be contrasted to the top panel of Figure 1 where there was nothing that distinguishes $x=40$ from other radial locations $x=80,120, 160$ in the time traces. The bottom panel of Figure~\ref{f:tau_dyn} for $A=2.0$ also shows that the evolution of ${\cal E}$ at $x=40$ is different from that at other locations. That is, ${\cal E}$ helps us identifying and differentiating the behaviour of PDFs at different spatial points.

\section{Results for vorticity with varying adiabaticity}
\label{sec:vort}
%Figure 5
%Figure vorticity time trace A=0.25, A=1.00, A=2.00
\begin{figure}[ht]
{\vspace{2mm}
\includegraphics[width=10cm, height=7cm]{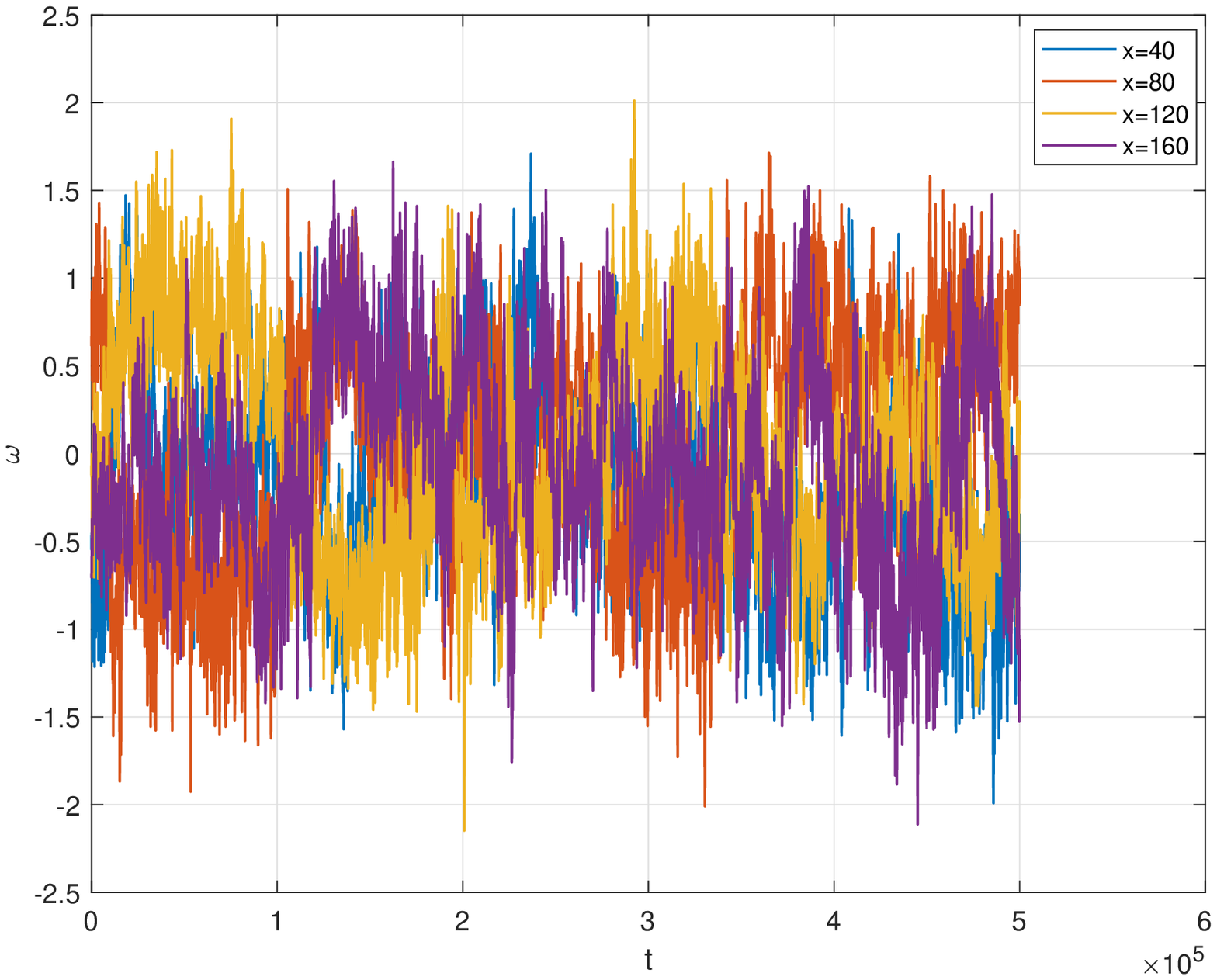}}
{\includegraphics[width=10cm, height=7cm]{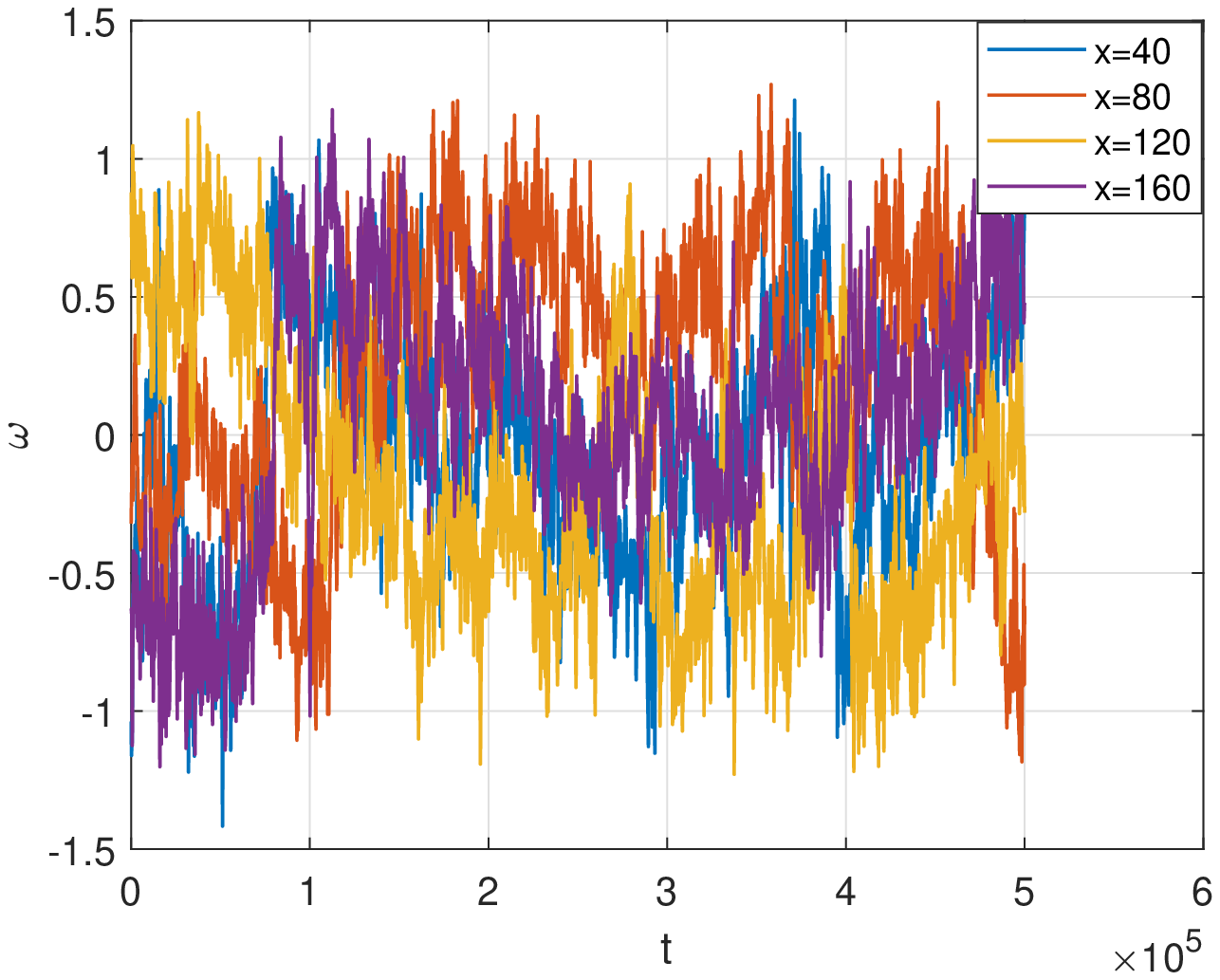}}
{\includegraphics[width=10cm, height=7cm]{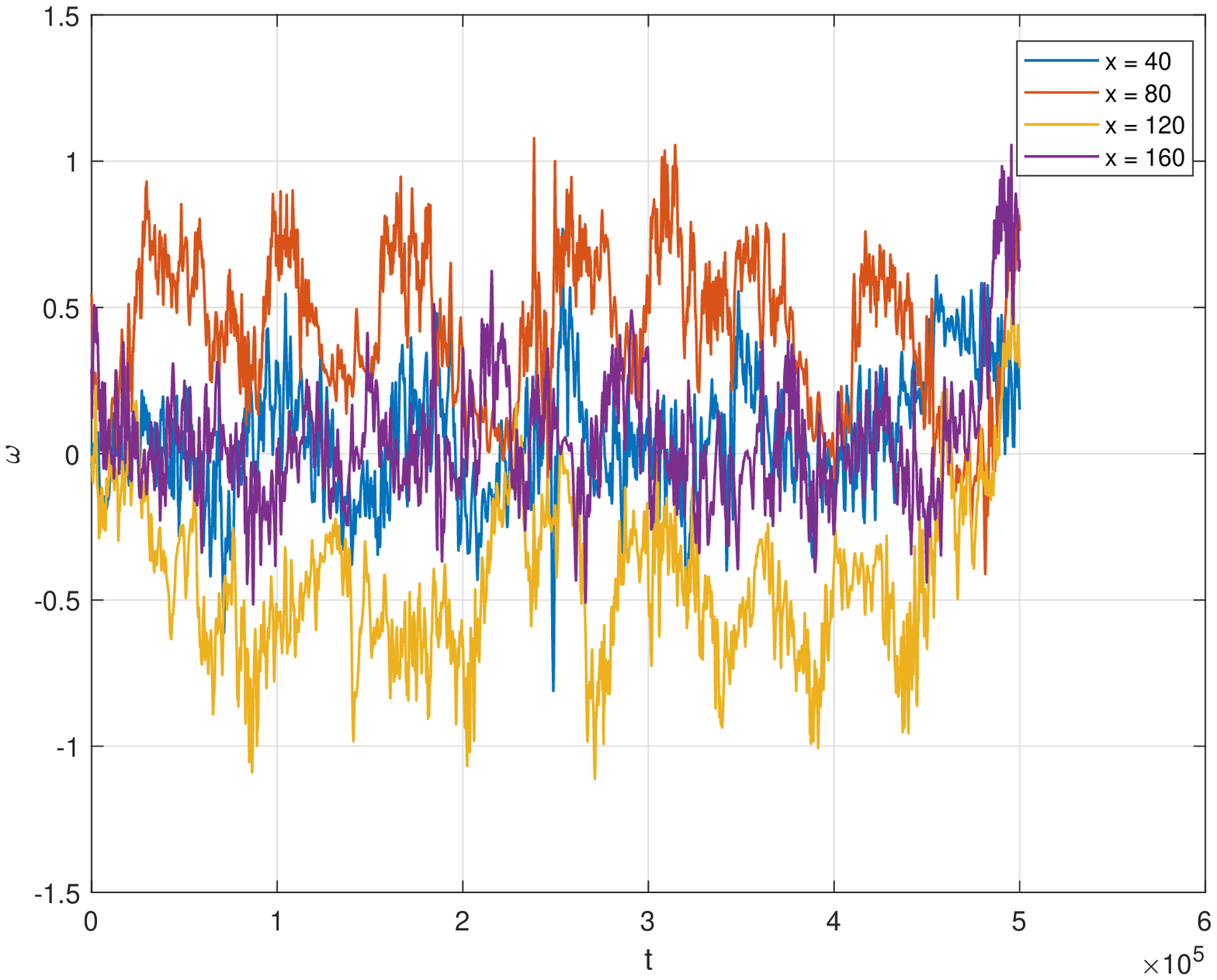}}
\caption{
The time trace of vorticity at the radial positions (40, 80, 120, 160) for $A = 0.25$ (top), $1.00$ (middle) and $2.00$ (bottom).
}
\label{f:vorticity_time_trace}
\end{figure}

%Figure 6
%Figure PDFs A=0.25
\begin{figure}[ht]
{\vspace{2mm}
\includegraphics[width=10cm, height=7cm]{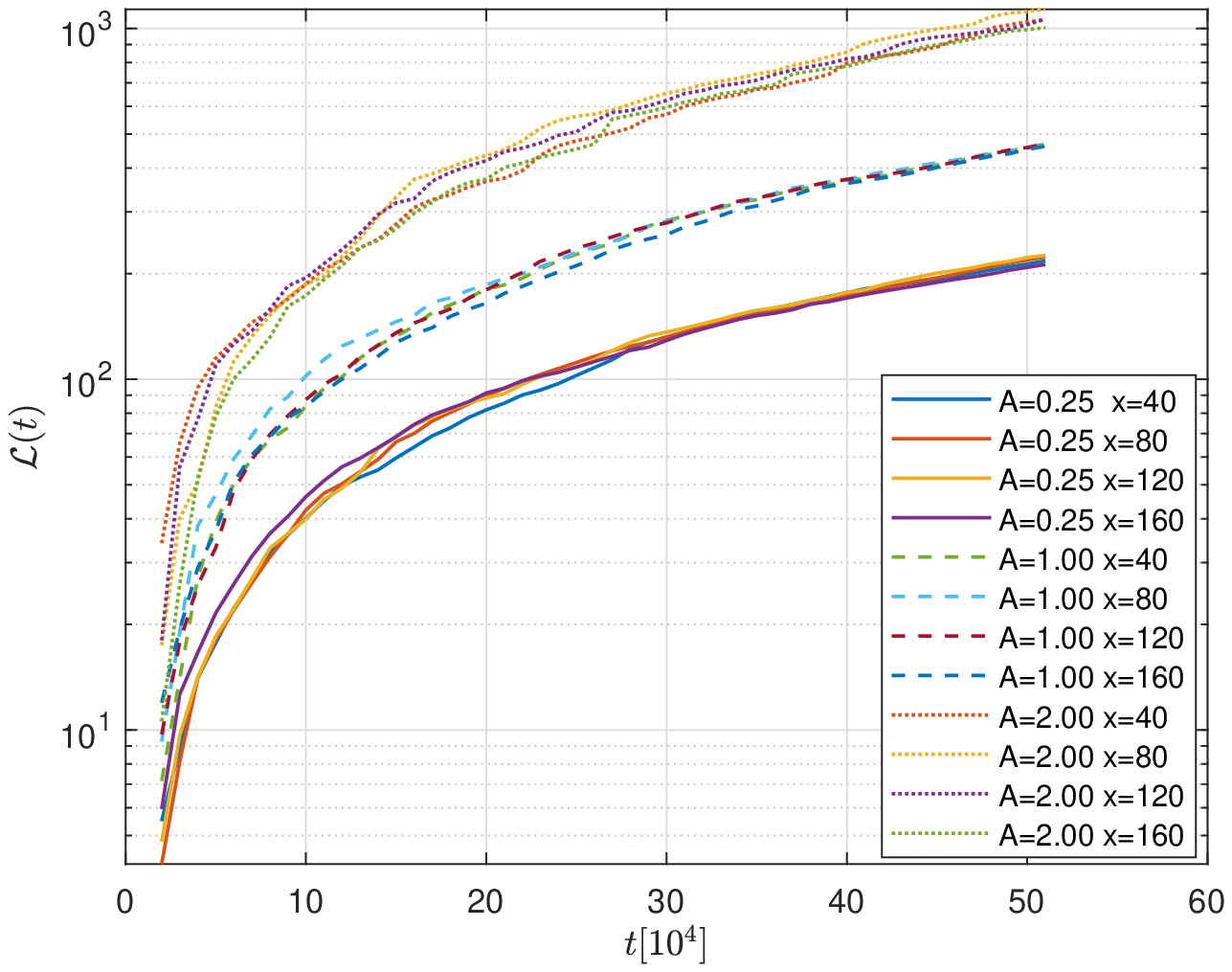}
}
\caption{
The logarithm of the information length of vorticity computed using Eq. (\ref{Int}), for the four radial points and $A = 0.25, 1.00$ and $2.00$.
}
\label{f:fsavort_int}
\end{figure}
In this section the results for vorticity is presented. First the time traces are shown in Figure \ref{f:vorticity_time_trace} and then in Figure \ref{f:fsavort_int} the information length is displayed. Overall, the time traces of vorticity exhibit similar behaviour to those of the potential where a short time scale oscillation is dominating for small adiabaticity whereas in addition to the short scale oscillations, an oscillation with a longer period emerges for $A=2.00$. This change in dynamics is also visible in the information length computed according to Eq. (\ref{Int}); the information length increases due to the change in dynamics. Moreover, the largest rate of change in information is in the early stages of the simulation, and the rate of change decreases as a system approaches a quasi-stationary state where the the variation in time-dependent PDFs is small. The slower variation on a longer time scale causes larger differences in the time evolution of the PDFs, yielding much larger values of information length although no sudden changes in the dynamics are visible.   
%%%%%%%%%%%%%%%%%%%%%%%%%%%%%%%%%%%%%%%%%%%%%%%%%%%%%%%%%%%%%%%%%%%%%%%%%%%%%%%%
\section{Results for flux with varying adiabaticity}
\label{sec:flux}
%Figure 7
%Figure vorticity time trace A=0.25, A=1.00, A=2.00
\begin{figure}[ht]
{\vspace{2mm}
\includegraphics[width=10cm, height=7cm]{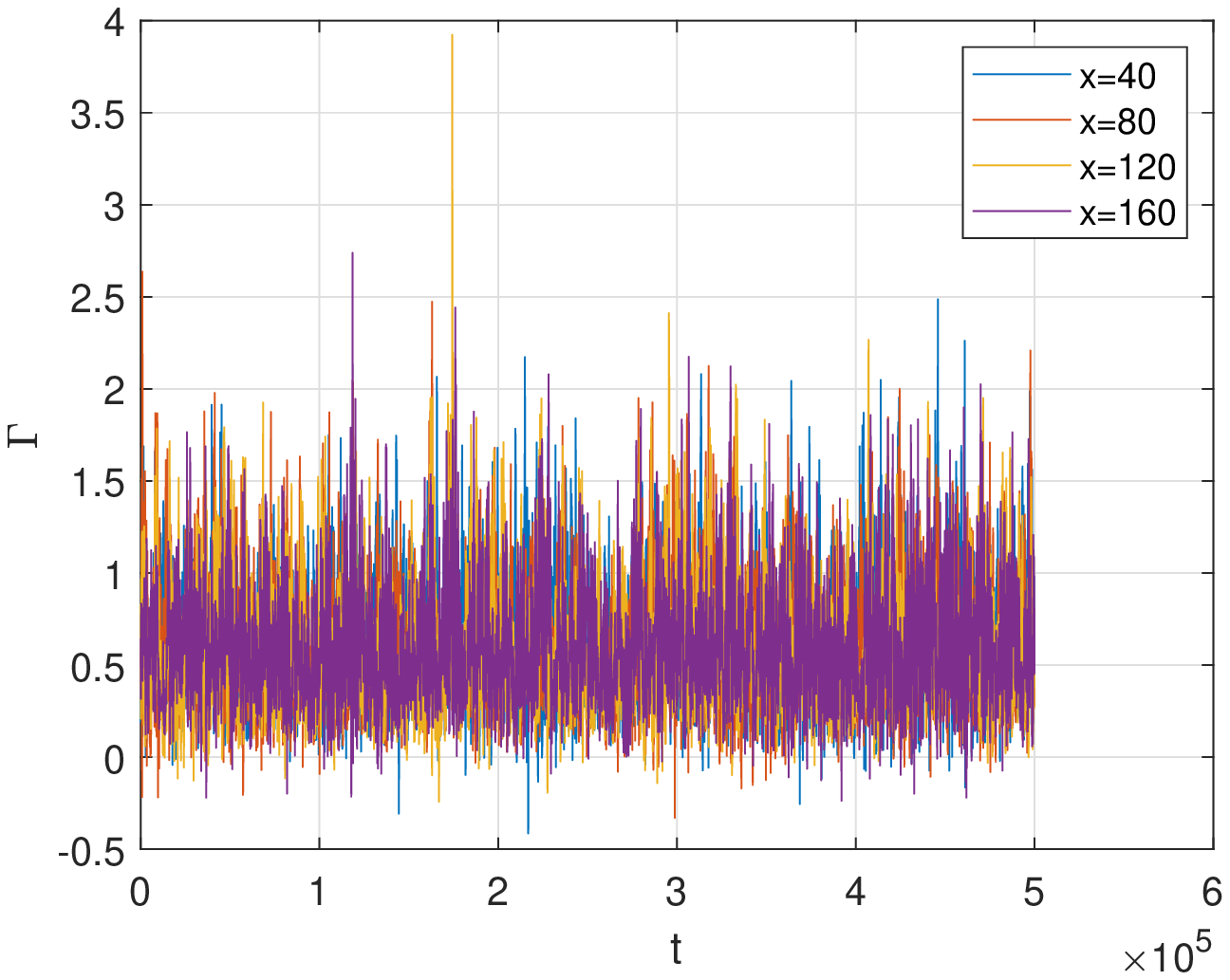}}
{\includegraphics[width=10cm, height=7cm]{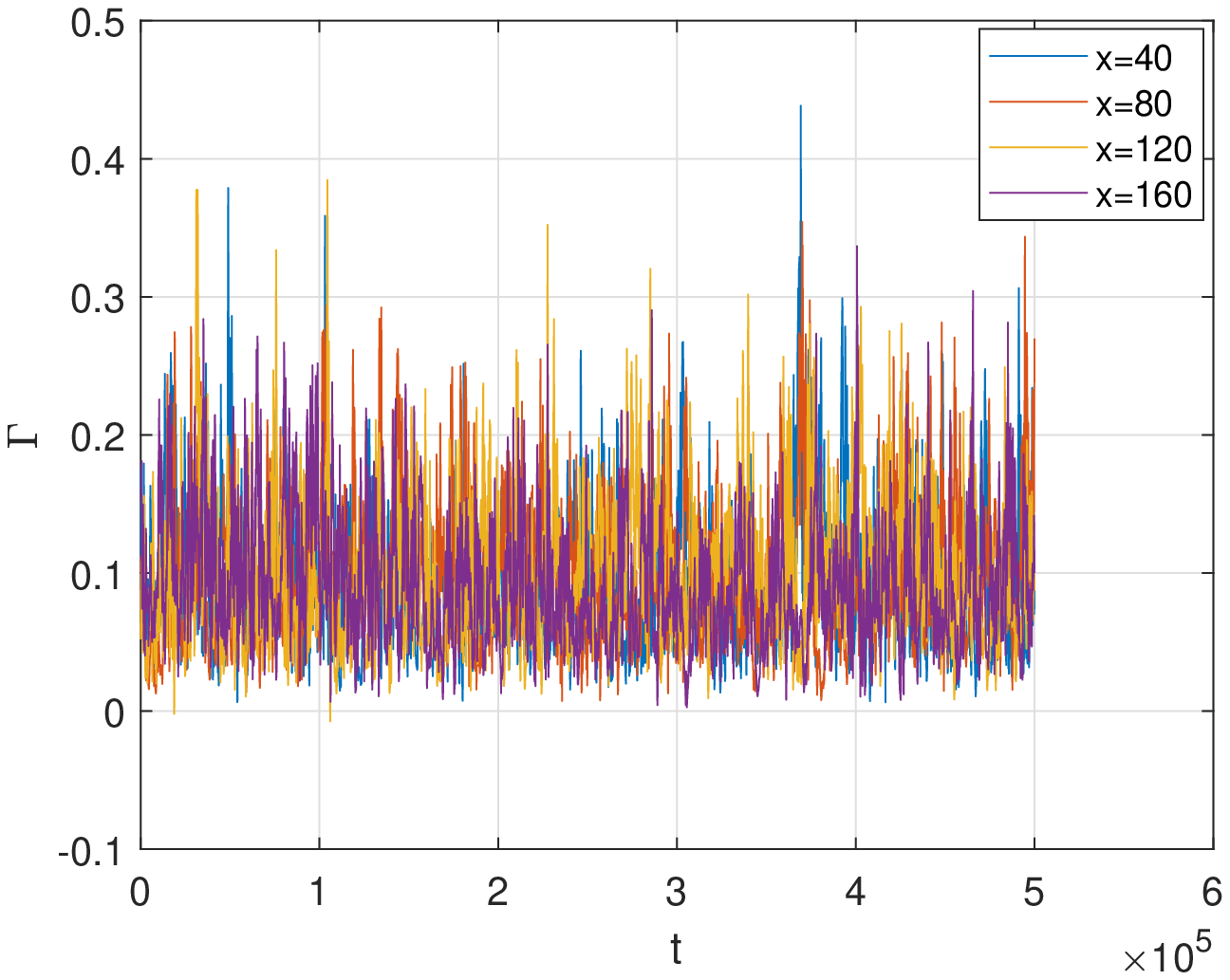}}
{\includegraphics[width=10cm, height=7cm]{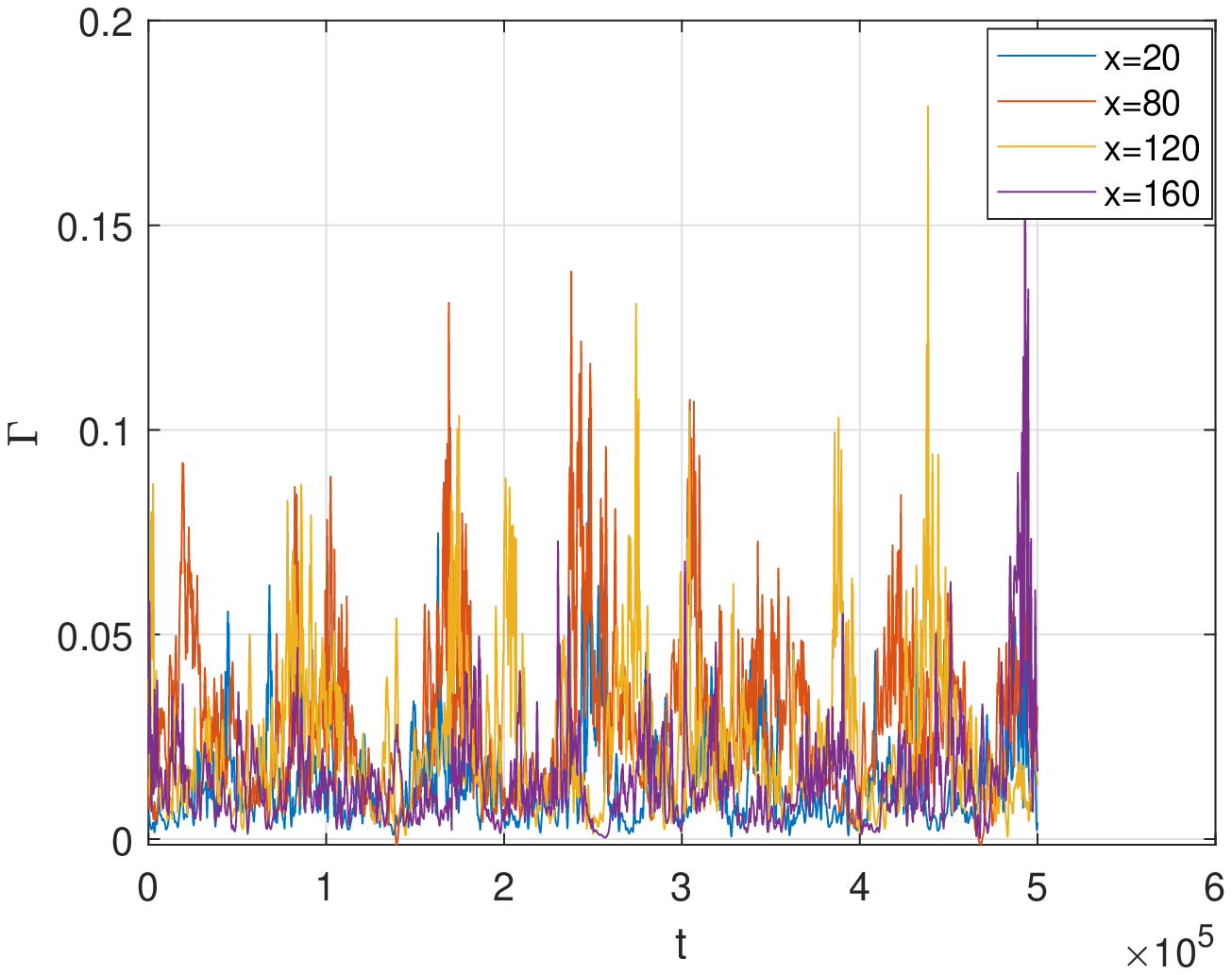}}
\caption{
The time trace of flux at the radial positions (40, 80, 120, 160) for $A = 0.25$ (top), $1.00$ (middle) and $2.00$ (bottom).
}
\label{f:time_trace_flux}
\end{figure}

%Figure 8
%Figure PDFs A=0.25
\begin{figure}[ht]
{\vspace{2mm}
\includegraphics[width=10cm, height=7cm]{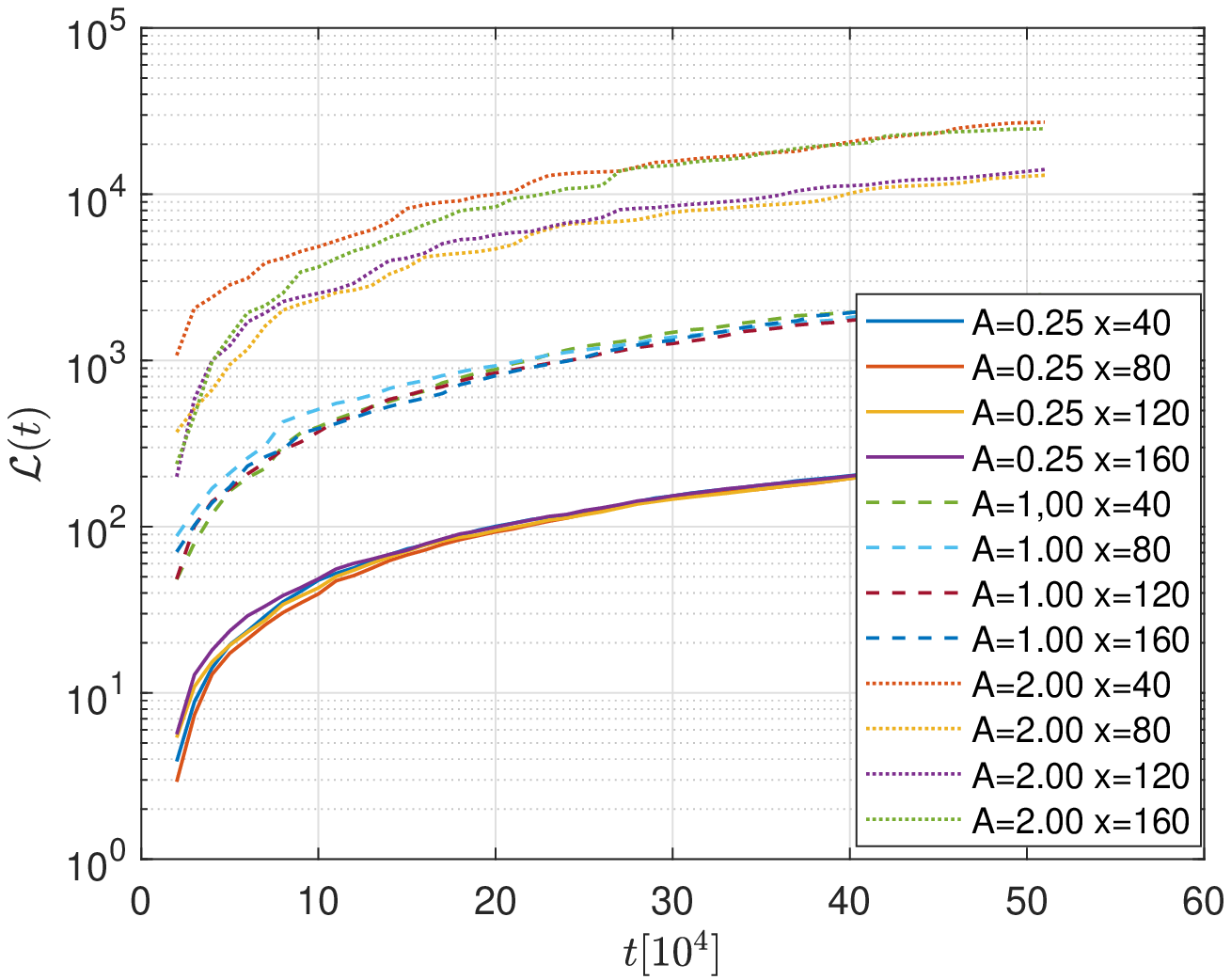}
}
\caption{
The logarithm of the information length of flux computed using Eq. (\ref{Int}), for the four radial points and $A = 0.25, 1.00$ and $2.00$.
}
\label{f:fsaflux_int}
\end{figure}
In this section we examine the poloidally averaged density flux, which is calculated by integral \eqref{fsa}:
\begin{equation}
 \Gamma=\langle \Gamma \rangle= \frac{1}{L_y} \int_0^{L_y} n \frac{\partial \phi}{\partial y} dy.
\label{fsaGamma}
\end{equation}
We note that Equation (\ref{fsaGamma}) described a second order quantity, i.e. an integral of a product of two fluctuating quantities which  thus have unique statistical features different to the statistics of the potential and the density. The three different regimes are elucidated comparing the time traces in Figure (\ref{f:time_trace_flux}), and the fast oscillation for small adiabaticity is present whereas the longer time scale fluctuation is visible for larger adiabaticity. Note that the fluctuation level is very different with larger fluctuations for smaller adiabaticity. Statistical features of flux time traces have been assessed and discussed previously in Ref. \onlinecite{anderson2017}. The information length is displayed in Figure (\ref{f:fsaflux_int}) where a large difference in information distance is visible in the early stages as well as two different regimes between the edge ($x=40$ and $x=160$) and core radial ($x=80$ and $x=120$) points. This difference is also visible in the time trace where there is a difference in fluctuation levels between the regimes.  
%%%%%%%%%%%%%%%%%%%%%%%%%%%%%%%%%%%%%%%%%%%%%%%%%%%%%%%%%%%%%%%%%%%%%%%%%%%%%%%%
\section{Discussion}
\label{discuss}
The physical understanding of turbulent transport processes have been developed from the Hasegawa - Mima (HM) and Hasegawa - Wakatani (HW) models that include qualitative features and scalings. However recently more accurate models have been considered such as gyrokinetics. In this work, we have performed a statistical analysis of the time traces for potential and vorticity generated by fluid like simulations of the HW model. The evolution of the time-dependent probability distribution functions (PDFs) have been analysed by a statistical measure called the information length. 
The dynamical time $\tau$ is a positive definite measure of the correlation time over which the (dimensionless) information changes. The information length represents the total different number of states between the initial and final times and establishes a distance between the initial and final PDFs in the statistical space.

Here it is worth noting that a signature of transport caused by coherent structures is elevated tails in comparison with the Gaussian distribution. It is found that the statistics of quantities describing transport mediated by coherent structures exhibist non-Gaussian features and have significantly increased kurtosis (Kurtosis is the normalized fourth moment, where a Gaussian process has a kurtosis of $3$.), for instance, as shown in the previous work in Ref. \onlinecite{anderson2017}.

From this perspective it is interesting to compare the information length found for the Gaussian process since the objective is to characterize the effect of generation and evolution of coherent structures. The information length or the dynamic time $\tau$ can be a useful diagnostics for accessing large events, or intermittent transport due to coherent structures. 

The time traces of the potential and density from the simulations are averaged in the poloidal direction. While we analyse the simulation data at ten radial positions however for simplicity, in the figures, the results for the positions $x= 40, 80, 120$ and $160$ only are shown. The simulation length is 500 001 time steps where the PDFs are constructed using $t=10 000$ time steps where one PDF per 100 000 time steps is shown. We observe a variation in mean and variance of the PDFs as time progresses, see e.g. Figures~\ref{f:fsaphi} and \ref{f:vorticity_time_trace}.  The expression in (\ref{etau}) is positive definite and thus the integral sum is not a decreasing function of time; see e.g. Figures~\ref{f:fsaphi_int}, \ref{f:fsavort_int} and \ref{f:fsaflux_int}. In fact, the integral could almost linearly increase with time but with large fluctuations. Note that the time trace of vorticity is varying faster than the potential, leading to a larger information length for the vorticity than for the potential. Moreover, there is a distinct difference between $A=0.25$ and $A=2.00$ cases due to the difference in dynamics, for $A=0$ the system reduces to the Euler equation. The oscillations (on a fast and a slower time scale) in time give rise to PDFs that are localized at different values of potential $\phi$ or vorticity $\omega$. Overall, a smaller $A$ leads to a smaller information length due to large fluctuation (uncertainty).
 
Here it is important to stress that this is a novel methodology of assessing the effects of coherent structures and turbulent dynamics in plasmas. The model and method is chosen for the reasons that it inherently includes coherent structures and qualitatively describes the complex plasma dynamics in magnetically confined plasmas. Moreover, it is feasible to generate a plasma fluid simulation containing many turnover times of the turbulent process such that the time traces can be expected to exhibit enough information for this type of analysis. The results suggest that this opens a new promising field of application of information length. The results suggest that it is feasible to find changes in the dynamics as a sudden change in distance between PDFs assessed by the information space, thus it could be a valuable tool for the investigation of turbulent plasma dynamics. 
\newpage
%\section*{References}

\end{document}